\documentclass[titlepage,fleqn]{article}
\usepackage{graphics}
\newcommand{\vertsinglea}{      %
\begin{minipage}{18mm}          %
\unitlength1mm                  %
\begin{picture}(18,16)          %
\thinlines                      %
\put( 0, 8){\line(1, 0){10}}    %
\put(10, 8){\line(1, 1){7}}     %
\put(10, 8){\line(1,-1){7}}     %
\put(10, 8){\circle*{3}}        %
\put( 3, 6){$\mu_1$}            %
\put(11,13){$\mu_3$}            %
\put(14, 5){$\mu_2$}            %
\end{picture}                   %
\end{minipage}}                 %

\newcommand{\vertfpair}{        %
\begin{minipage}{26mm}          %
\unitlength1mm                  %
\begin{picture}(26,26)          %
\thinlines                      %
\put( 8,13){\line(-1, 1){7}}    %
\put( 8,13){\line(-1,-1){7}}    %
\put( 8,13){\line( 1, 0){10}}   %
\put(18,13){\line( 1, 1){7}}    %
\put(18,13){\line( 1,-1){7}}    %
\put( 8,13){\circle*{3}}        %
\put(18,13){\circle*{3}}        %
\put( 0,22){$\mu_1$}            %
\put(22,22){$\mu_3$}            %
\put(22, 3){$\mu_4$}            %
\put( 0, 3){$\mu_2$}            %
\end{picture}                   %
\end{minipage}}                 %

\newcommand{\vertfpairij}{      %
\begin{minipage}{26mm}          %
\unitlength1mm                  %
\begin{picture}(26,26)          %
\thinlines                      %
\put( 8,13){\line(-1, 1){7}}    %
\put( 8,13){\line(-1,-1){7}}    %
\put( 8,13){\line( 1, 0){10}}   %
\put(18,13){\line( 1, 1){7}}    %
\put(18,13){\line( 1,-1){7}}    %
\put( 8,13){\circle*{3}}        %
\put(18,13){\circle*{3}}        %
\put( 0,22){$\mu_1$}            %
\put(22,22){$\mu_3$}            %
\put(22, 3){$\mu_4$}            %
\put( 0, 3){$\mu_2$}            %
\put( 7, 8){$i$}                %
\put(17, 8){$j$}                %
\end{picture}                   %
\end{minipage}}                 %

\newcommand{\vertlii}{          %
\begin{minipage}{18mm}          %
\unitlength1mm                  %
\begin{picture}(18,16)          %
\thinlines                      %
\put( 0, 8){\line(1, 0){10}}    %
\put(10, 8){\line(1, 1){7}}     %
\put(10, 8){\line(1,-1){7}}     %
\put(10, 8){\circle*{3}}        %
\put( 3, 8){\line(2, 1){4}}     %
\put( 3, 8){\line(2,-1){4}}     %
\put(12,10){\line(1, 2){2}}     %
\put(12,10){\line(2, 1){4}}     %
\put(12, 6){\line(1,-2){2}}     %
\put(12, 6){\line(2,-1){4}}     %
\end{picture}                   %
\end{minipage}}                 %

\newcommand{\vertrii}{          %
\begin{minipage}{18mm}          %
\unitlength1mm                  %
\begin{picture}(18,16)          %
\thinlines                      %
\put( 0, 8){\line(1, 0){10}}    %
\put(10, 8){\line(1, 1){7}}     %
\put(10, 8){\line(1,-1){7}}     %
\put(10, 8){\circle*{3}}        %
\put( 7, 8){\line(-2, 1){4}}    %
\put( 7, 8){\line(-2,-1){4}}    %
\put(12,10){\line( 1, 2){2}}    %
\put(12,10){\line( 2, 1){4}}    %
\put(12, 6){\line( 1,-2){2}}    %
\put(12, 6){\line( 2,-1){4}}    %
\end{picture}                   %
\end{minipage}}                 %

\newcommand{\vertloi}{          %
\begin{minipage}{18mm}          %
\unitlength1mm                  %
\begin{picture}(18,16)          %
\thinlines                      %
\put( 0, 8){\line(1, 0){10}}    %
\put(10, 8){\line(1, 1){7}}     %
\put(10, 8){\line(1,-1){7}}     %
\put(10, 8){\circle*{3}}        %
\put( 3, 8){\line( 2, 1){4}}    %
\put( 3, 8){\line( 2,-1){4}}    %
\put(12,10){\line( 1, 2){2}}    %
\put(12,10){\line( 2, 1){4}}    %
\put(16, 2){\line(-1, 2){2}}    %
\put(16, 2){\line(-2, 1){4}}    %
\end{picture}                   %
\end{minipage}}                 %

\newcommand{\vertroi}{          %
\begin{minipage}{18mm}          %
\unitlength1mm                  %
\begin{picture}(18,16)          %
\thinlines                      %
\put( 0, 8){\line(1, 0){10}}    %
\put(10, 8){\line(1, 1){7}}     %
\put(10, 8){\line(1,-1){7}}     %
\put(10, 8){\circle*{3}}        %
\put( 7, 8){\line(-2, 1){4}}    %
\put( 7, 8){\line(-2,-1){4}}    %
\put(12,10){\line( 1, 2){2}}    %
\put(12,10){\line( 2, 1){4}}    %
\put(16, 2){\line(-1, 2){2}}    %
\put(16, 2){\line(-2, 1){4}}    %
\end{picture}                   %
\end{minipage}}                 %

\newcommand{\vertlio}{          %
\begin{minipage}{18mm}          %
\unitlength1mm                  %
\begin{picture}(18,16)          %
\thinlines                      %
\put( 0, 8){\line(1, 0){10}}    %
\put(10, 8){\line(1, 1){7}}     %
\put(10, 8){\line(1,-1){7}}     %
\put(10, 8){\circle*{3}}        %
\put( 3, 8){\line( 2, 1){4}}    %
\put( 3, 8){\line( 2,-1){4}}    %
\put(16,14){\line(-1,-2){2}}    %
\put(16,14){\line(-2,-1){4}}    %
\put(12, 6){\line( 1,-2){2}}    %
\put(12, 6){\line( 2,-1){4}}    %
\end{picture}                   %
\end{minipage}}                 %

\newcommand{\vertrio}{          %
\begin{minipage}{18mm}          %
\unitlength1mm                  %
\begin{picture}(18,16)          %
\thinlines                      %
\put( 0, 8){\line(1, 0){10}}    %
\put(10, 8){\line(1, 1){7}}     %
\put(10, 8){\line(1,-1){7}}     %
\put(10, 8){\circle*{3}}        %
\put( 7, 8){\line(-2, 1){4}}    %
\put( 7, 8){\line(-2,-1){4}}    %
\put(16,14){\line(-1,-2){2}}    %
\put(16,14){\line(-2,-1){4}}    %
\put(12, 6){\line( 1,-2){2}}    %
\put(12, 6){\line( 2,-1){4}}    %
\end{picture}                   %
\end{minipage}}                 %

\newcommand{\vertloo}{          %
\begin{minipage}{18mm}          %
\unitlength1mm                  %
\begin{picture}(18,16)          %
\thinlines                      %
\put( 0, 8){\line(1, 0){10}}    %
\put(10, 8){\line(1, 1){7}}     %
\put(10, 8){\line(1,-1){7}}     %
\put(10, 8){\circle*{3}}        %
\put( 3, 8){\line( 2, 1){4}}    %
\put( 3, 8){\line( 2,-1){4}}    %
\put(16,14){\line(-1,-2){2}}    %
\put(16,14){\line(-2,-1){4}}    %
\put(16, 2){\line(-1, 2){2}}    %
\put(16, 2){\line(-2, 1){4}}    %
\end{picture}                   %
\end{minipage}}                 %

\newcommand{\vertroo}{          %
\begin{minipage}{18mm}          %
\unitlength1mm                  %
\begin{picture}(18,16)          %
\thinlines                      %
\put( 0, 8){\line(1, 0){10}}    %
\put(10, 8){\line(1, 1){7}}     %
\put(10, 8){\line(1,-1){7}}     %
\put(10, 8){\circle*{3}}        %
\put( 7, 8){\line(-2, 1){4}}    %
\put( 7, 8){\line(-2,-1){4}}    %
\put(16,14){\line(-1,-2){2}}    %
\put(16,14){\line(-2,-1){4}}    %
\put(16, 2){\line(-1, 2){2}}    %
\put(16, 2){\line(-2, 1){4}}    %
\end{picture}                   %
\end{minipage}}                 %

\newcommand{\vertiil}{          %
\begin{minipage}{18mm}          %
\unitlength1mm                  %
\begin{picture}(18,16)          %
\thinlines                      %
\put( 8, 8){\line( 1, 0){10}}   %
\put( 8, 8){\line(-1, 1){7}}    %
\put( 8, 8){\line(-1,-1){7}}    %
\put( 8, 8){\circle*{3}}        %
\put( 6, 6){\line(-2,-1){4}}    %
\put( 6, 6){\line(-1,-2){2}}    %
\put( 6,10){\line(-1, 2){2}}    %
\put( 6,10){\line(-2, 1){4}}    %
\put(11, 8){\line( 2, 1){4}}    %
\put(11, 8){\line( 2,-1){4}}    %
\end{picture}                   %
\end{minipage}}                 %

\newcommand{\vertiir}{          %
\begin{minipage}{18mm}          %
\unitlength1mm                  %
\begin{picture}(18,16)          %
\thinlines                      %
\put( 8, 8){\line( 1, 0){10}}   %
\put( 8, 8){\line(-1, 1){7}}    %
\put( 8, 8){\line(-1,-1){7}}    %
\put( 8, 8){\circle*{3}}        %
\put( 6, 6){\line(-2,-1){4}}    %
\put( 6, 6){\line(-1,-2){2}}    %
\put( 6,10){\line(-1, 2){2}}    %
\put( 6,10){\line(-2, 1){4}}    %
\put(15, 8){\line(-2, 1){4}}    %
\put(15, 8){\line(-2,-1){4}}    %
\end{picture}                   %
\end{minipage}}                 %

\newcommand{\vertiol}{          %
\begin{minipage}{18mm}          %
\unitlength1mm                  %
\begin{picture}(18,16)          %
\thinlines                      %
\put( 8, 8){\line( 1, 0){10}}   %
\put( 8, 8){\line(-1, 1){7}}    %
\put( 8, 8){\line(-1,-1){7}}    %
\put( 8, 8){\circle*{3}}        %
\put( 2, 2){\line( 2, 1){4}}    %
\put( 2, 2){\line( 1, 2){2}}    %
\put( 6,10){\line(-1, 2){2}}    %
\put( 6,10){\line(-2, 1){4}}    %
\put(11, 8){\line( 2, 1){4}}    %
\put(11, 8){\line( 2,-1){4}}    %
\end{picture}                   %
\end{minipage}}                 %

\newcommand{\vertior}{          %
\begin{minipage}{18mm}          %
\unitlength1mm                  %
\begin{picture}(18,16)          %
\thinlines                      %
\put( 8, 8){\line( 1, 0){10}}   %
\put( 8, 8){\line(-1, 1){7}}    %
\put( 8, 8){\line(-1,-1){7}}    %
\put( 8, 8){\circle*{3}}        %
\put( 2, 2){\line( 2, 1){4}}    %
\put( 2, 2){\line( 1, 2){2}}    %
\put( 6,10){\line(-1, 2){2}}    %
\put( 6,10){\line(-2, 1){4}}    %
\put(15, 8){\line(-2, 1){4}}    %
\put(15, 8){\line(-2,-1){4}}    %
\end{picture}                   %
\end{minipage}}                 %

\newcommand{\vertoil}{          %
\begin{minipage}{18mm}          %
\unitlength1mm                  %
\begin{picture}(18,16)          %
\thinlines                      %
\put( 8, 8){\line( 1, 0){10}}   %
\put( 8, 8){\line(-1, 1){7}}    %
\put( 8, 8){\line(-1,-1){7}}    %
\put( 8, 8){\circle*{3}}        %
\put( 6, 6){\line(-2,-1){4}}    %
\put( 6, 6){\line(-1,-2){2}}    %
\put( 2,14){\line( 1,-2){2}}    %
\put( 2,14){\line( 2,-1){4}}    %
\put(11, 8){\line( 2, 1){4}}    %
\put(11, 8){\line( 2,-1){4}}    %
\end{picture}                   %
\end{minipage}}                 %

\newcommand{\vertoir}{          %
\begin{minipage}{18mm}          %
\unitlength1mm                  %
\begin{picture}(18,16)          %
\thinlines                      %
\put( 8, 8){\line( 1, 0){10}}   %
\put( 8, 8){\line(-1, 1){7}}    %
\put( 8, 8){\line(-1,-1){7}}    %
\put( 8, 8){\circle*{3}}        %
\put( 6, 6){\line(-2,-1){4}}    %
\put( 6, 6){\line(-1,-2){2}}    %
\put( 2,14){\line( 1,-2){2}}    %
\put( 2,14){\line( 2,-1){4}}    %
\put(15, 8){\line(-2, 1){4}}    %
\put(15, 8){\line(-2,-1){4}}    %
\end{picture}                   %
\end{minipage}}                 %

\newcommand{\vertool}{          %
\begin{minipage}{18mm}          %
\unitlength1mm                  %
\begin{picture}(18,16)          %
\thinlines                      %
\put( 8, 8){\line( 1, 0){10}}   %
\put( 8, 8){\line(-1, 1){7}}    %
\put( 8, 8){\line(-1,-1){7}}    %
\put( 8, 8){\circle*{3}}        %
\put( 2, 2){\line( 2, 1){4}}    %
\put( 2, 2){\line( 1, 2){2}}    %
\put( 2,14){\line( 1,-2){2}}    %
\put( 2,14){\line( 2,-1){4}}    %
\put(11, 8){\line( 2, 1){4}}    %
\put(11, 8){\line( 2,-1){4}}    %
\end{picture}                   %
\end{minipage}}                 %

\newcommand{\vertoor}{          %
\begin{minipage}{18mm}          %
\unitlength1mm                  %
\begin{picture}(18,16)          %
\thinlines                      %
\put( 8, 8){\line( 1, 0){10}}   %
\put( 8, 8){\line(-1, 1){7}}    %
\put( 8, 8){\line(-1,-1){7}}    %
\put( 8, 8){\circle*{3}}        %
\put( 2, 2){\line( 2, 1){4}}    %
\put( 2, 2){\line( 1, 2){2}}    %
\put( 2,14){\line( 1,-2){2}}    %
\put( 2,14){\line( 2,-1){4}}    %
\put(15, 8){\line(-2, 1){4}}    %
\put(15, 8){\line(-2,-1){4}}    %
\end{picture}                   %
\end{minipage}}                 %

\newcommand{\vertxlii}{         %
\begin{minipage}{18mm}          %
\unitlength1mm                  %
\begin{picture}(18,16)          %
\thicklines                     %
\put(10, 8){\circle{3}}         %
\put( 0, 8){\line(1, 0){4}}     %
\put(10, 8){\vector(-1, 0){6}}  %
\put(10, 8){\line(1, 1){4}}     %
\put(17,15){\vector(-1,-1){4}}  %
\put(10, 8){\line(1,-1){4}}     %
\put(17, 1){\vector(-1, 1){4}}  %
\end{picture}                   %
\end{minipage}}                 %

\newcommand{\vertxrii}{         %
\begin{minipage}{18mm}          %
\unitlength1mm                  %
\begin{picture}(18,16)          %
\thicklines                     %
\put(10, 8){\circle{3}}         %
\put( 6, 8){\line(1, 0){4}}     %
\put( 0, 8){\vector( 1, 0){6}}  %
\put(10, 8){\line(1, 1){4}}     %
\put(17,15){\vector(-1,-1){4}}  %
\put(10, 8){\line(1,-1){4}}     %
\put(17, 1){\vector(-1, 1){4}}  %
\end{picture}                   %
\end{minipage}}                 %

\newcommand{\vertxloi}{         %
\begin{minipage}{18mm}          %
\unitlength1mm                  %
\begin{picture}(18,16)          %
\thicklines                     %
\put(10, 8){\circle{3}}         %
\put( 0, 8){\line(1, 0){4}}     %
\put(10, 8){\vector(-1, 0){6}}  %
\put(10, 8){\line(1, 1){4}}     %
\put(17,15){\vector(-1,-1){4}}  %
\put(17, 1){\line(-1,1){4}}     %
\put(10, 8){\vector( 1,-1){5}}  %
\end{picture}                   %
\end{minipage}}                 %

\newcommand{\vertxroi}{         %
\begin{minipage}{18mm}          %
\unitlength1mm                  %
\begin{picture}(18,16)          %
\thicklines                     %
\put(10, 8){\circle{3}}         %
\put( 6, 8){\line(1, 0){4}}     %
\put( 0, 8){\vector( 1, 0){6}}  %
\put(10, 8){\line(1, 1){4}}     %
\put(17,15){\vector(-1,-1){4}}  %
\put(17, 1){\line(-1,1){4}}     %
\put(10, 8){\vector( 1,-1){5}}  %
\end{picture}                   %
\end{minipage}}                 %

\newcommand{\vertxlio}{         %
\begin{minipage}{18mm}          %
\unitlength1mm                  %
\begin{picture}(18,16)          %
\thicklines                     %
\put(10, 8){\circle{3}}         %
\put( 0, 8){\line( 1, 0){4}}    %
\put(10, 8){\vector(-1, 0){6}}  %
\put(17,15){\line(-1,-1){4}}    %
\put(10, 8){\vector( 1, 1){5}}  %
\put(10, 8){\line( 1,-1){4}}    %
\put(17, 1){\vector(-1, 1){4}}  %
\end{picture}                   %
\end{minipage}}                 %

\newcommand{\vertxrio}{         %
\begin{minipage}{18mm}          %
\unitlength1mm                  %
\begin{picture}(18,16)          %
\thicklines                     %
\put(10, 8){\circle{3}}         %
\put( 6, 8){\line(1, 0){4}}     %
\put( 0, 8){\vector( 1, 0){6}}  %
\put(17,15){\line(-1,-1){4}}    %
\put(10, 8){\vector( 1, 1){5}}  %
\put(10, 8){\line( 1,-1){4}}    %
\put(17, 1){\vector(-1, 1){4}}  %
\end{picture}                   %
\end{minipage}}                 %

\newcommand{\vertxloo}{         %
\begin{minipage}{18mm}          %
\unitlength1mm                  %
\begin{picture}(18,16)          %
\thicklines                     %
\put(10, 8){\circle{3}}         %
\put( 0, 8){\line( 1, 0){4}}    %
\put(10, 8){\vector(-1, 0){6}}  %
\put(17,15){\line(-1,-1){4}}    %
\put(10, 8){\vector( 1, 1){5}}  %
\put(17, 1){\line(-1, 1){4}}    %
\put(10, 8){\vector( 1,-1){5}}  %
\end{picture}                   %
\end{minipage}}                 %

\newcommand{\vertxroo}{         %
\begin{minipage}{18mm}          %
\unitlength1mm                  %
\begin{picture}(18,16)          %
\thicklines                     %
\put(10, 8){\circle{3}}         %
\put( 6, 8){\line(1, 0){4}}     %
\put( 0, 8){\vector( 1, 0){6}}  %
\put(17,15){\line(-1,-1){4}}    %
\put(10, 8){\vector( 1, 1){5}}  %
\put(17, 1){\line(-1, 1){4}}    %
\put(10, 8){\vector( 1,-1){5}}  %
\end{picture}                   %
\end{minipage}}                 %

\newcommand{\zp}[1]{#1}
\newcommand{\zn}[1]{\overline #1}
\newcommand{\neel}{N\'{e}el}
\renewcommand{\phi}{\varphi}
\renewcommand{\rho}{\varrho}

\begin{document}
\begin{titlepage}
\renewcommand{\thefootnote}{\ensuremath{\fnsymbol{footnote}}}
\begin{center}
\Large
Quantum phase transition in spin-$\frac{3}{2}$ systems on the
hexagonal lattice -- optimum ground state approach\footnote{Work
performed within the research program of the
Sonderforschungsbereich 341, K\"{o}ln-Aachen-J\"{u}lich}
\end{center}

\vspace{0.3cm}

\begin{center}
\large
H.~Niggemann \hspace{1cm} A.~Kl\"{u}mper \hspace{1cm} J.~Zittartz
\end{center}

\vspace{0.3cm}

\begin{center}
\small
Institut f\"{u}r Theoretische Physik, Universit\"{a}t zu K\"{o}ln, \newline
Z\"{u}lpicher Strasse 77, D-50937 K\"{o}ln, Germany \newline
(email: hn@thp.uni-koeln.de, kluemper@thp.uni-koeln.de, zitt@thp.uni-koeln.de)
\end{center}

\vspace{1cm}

{\small
Optimum ground states are constructed in two dimensions by using so
called {\em vertex state models}. These models are graphical generalizations of
the well-known matrix product ground states for spin chains. On the hexagonal
lattice we obtain
a one-parametric set of ground states for a five-dimensional manifold of
$S=\frac{3}{2}$ Hamiltonians. Correlation functions within these ground states
are calculated using Monte-Carlo simulations. In contrast to the one-dimensional
situation, these states exhibit a parameter-induced second order phase transition.
In the disordered phase, two-spin correlations decay exponentially, but in the
\neel\ ordered phase alternating long-range correlations are dominant.
We also show that ground state properties can be obtained from the exact
solution of a corresponding free-fermion model for most values of the
parameter.
}
\end{titlepage}
\setcounter{footnote}{0}
\renewcommand{\thefootnote}{\arabic{footnote}}
\section{Introduction}
\label{sec_intro}
In recent work [1-4], we have presented optimum ground states for quantum mechanial
spin-1 and spin-$\frac{3}{2}$ chains with nearest neighbour interaction.
Optimum ground states are exact ground states which are simultaneously ground
states of the local interaction. They do not involve any excited local states.
These optimum ground states were constructed using multiple products of
finite matrices. In \cite{nz} we also presented an alternative graphical
representation of the constructed state, which we called {\em vertex state model}.
This representation has the advantage that it can be easily generalized to
higher dimensional spin systems.

In the present work we apply the concept of vertex state models to construct optimum
ground states for spin-$\frac{3}{2}$ on the hexagonal lattice. The set of
global ground states is parametrized by an anisotropy parameter, which determines
most properties of the ground state. In contrast to the optimum ground states
for spin chains, this two-dimensional model exhibits a parameter controlled
second order phase transition.

The paper is organized as follows: In section \ref{sec_32ham} we present
a general parametrization of all near\-est-neigh\-bour Hamiltonians for $S=\frac{3}{2}$
that obey a certain set of minimum symmetries. Section \ref{sec_constr}
contains a detailed explanation, how non-trivial optimum ground states can be
constructed for these Hamiltonians using vertex state models.
The properties of the constructed set of optimum ground states are
investigated in section \ref{sec_prop} using Monte-Carlo simulations.
The most important result is the existence of a parameter-induced second order
transition from a disorded to a \neel\ ordered phase. In section \ref{sec_free}
we show that ground state properties are asymptotically given by the solution
of a free-fermion model. The exact solution of this model confirms our
numerical results with high precision.
We finally give a short summary of our results in section \ref{sec_concl}.
\section{Spin-$\frac{3}{2}$ Hamiltonians}
\label{sec_32ham}
We shall construct optimum ground states for spin-$\frac{3}{2}$ on the
hexagonal lattice with periodic boundary conditions. In our previous paper
\cite{nz} we have already presented the most general Hamiltonian with nearest
neighbour interaction
\begin{equation}
\label{frm_globham}
H = \sum_{\langle i,j \rangle} h_{i,j}
\end{equation}
for $S=\frac{3}{2}$ systems with the following set of minimum symmetries:
\begin{itemize}
\item Homogeneity: All local Hamiltonians $h_{i,j}$ are equal, they only
      act on different spin pairs.
\item Parity invariance: $h_{i,j}$ commutes with the parity operator
      $P_{i,j}$, which interchanges the two neighbour spins $i$ and $j$.
\item Rotational invariance in the xy-plane of spin space: $h_{i,j}$
      commutes with the pair magnetization operator $S_i^z + S_j^z$.
\item Spin flip invariance: $S^z \rightarrow -S^z$ leaves $h_{i,j}$
      unchanged.
\end{itemize}
It turns out that all allowed local Hamiltonians can be written as
\begin{equation}
\begin{array}{rcl}
\label{frm_genham}
h_{i,j}& = & \lambda_3      (\,|v_3      \rangle\langle v_3      |    + 
                               |v_{-3}   \rangle\langle v_{-3}   |\,) + \\
       &   & \lambda_2^+    (\,|v_2^+    \rangle\langle v_2^+    |    +
                               |v_{-2}^+ \rangle\langle v_{-2}^+ |\,) + \\
       &   & \lambda_2^-    (\,|v_2^-    \rangle\langle v_2^-    |    +
                               |v_{-2}^- \rangle\langle v_{-2}^- |\,) + \\
       &   & \lambda_{11}^+ (\,|v_{11}^+ \rangle\langle v_{11}^+ |    +
                               |v_{-11}^+\rangle\langle v_{-11}^+|\,) + \\
       &   & \lambda_{12}^+ (\,|v_{12}^+ \rangle\langle v_{12}^+ |    +
                               |v_{-12}^+\rangle\langle v_{-12}^+|\,) + \\
       &   & \lambda_1^-    (\,|v_1^-    \rangle\langle v_1^-    |    +
                               |v_{-1}^- \rangle\langle v_{-1}^- |\,) + \\
       &   & \lambda_{01}^+    |v_{01}^+ \rangle\langle v_{01}^+ |    +
             \lambda_{02}^+    |v_{02}^+ \rangle\langle v_{02}^+ |    + \\
       &   & \lambda_{01}^-    |v_{01}^- \rangle\langle v_{01}^- |    +
             \lambda_{02}^-    |v_{02}^- \rangle\langle v_{02}^- | \; .
\end{array}
\end{equation}
The local states $v_{\mu,n}^p$ are defined in Table \ref{tab_32eigen},
where we have used the following notation for the canonical spin-$\frac{3}{2}$ basis
\begin{equation} \begin{array}{rclcrcl}
S_i^z\; |\zp{3}\rangle & = & \hphantom{-}\frac{3}{2} |\zp{3}\rangle & \hspace{1cm} &
S_i^z\; |\zp{1}\rangle & = & \hphantom{-}\frac{1}{2} |\zp{1}\rangle \\
S_i^z\; |\zn{3}\rangle & = & -\frac{3}{2} |\zn{3}\rangle & \hspace{1cm} &
S_i^z\; |\zn{1}\rangle & = & -\frac{1}{2} |\zn{1}\rangle \; .
\end{array} \end{equation}
In Table \ref{tab_32eigen}, $\mu$ is the pair magnetization, i.e. the eigenvalue of
$S_i^z + S_j^z$ and $p$ is the eigenvalue of the parity operator $P_{i,j}$.
\begin{table}[htb]
\caption{Eigenstates of the local Hamiltonian $h$}
\label{tab_32eigen}
\begin{center}
\begin{tabular}{|r|r|l|l|}
\hline
$\mu$ & $p$ & name & state \\
\hline
$ 3$ & $ 1$ & $v_3      $ & $|\zp{3}\zp{3}\rangle$ \\
$-3$ & $ 1$ & $v_{-3}   $ & $|\zn{3}\zn{3}\rangle$ \\
$ 2$ & $ 1$ & $v_2^+    $ & $|\zp{3}\zp{1}\rangle+|\zp{1}\zp{3}\rangle$ \\
$ 2$ & $-1$ & $v_2^-    $ & $|\zp{3}\zp{1}\rangle-|\zp{1}\zp{3}\rangle$ \\
$-2$ & $ 1$ & $v_{-2}^+ $ & $|\zn{3}\zn{1}\rangle+|\zn{1}\zn{3}\rangle$ \\
$-2$ & $-1$ & $v_{-2}^- $ & $|\zn{3}\zn{1}\rangle-|\zn{1}\zn{3}\rangle$ \\
$ 1$ & $ 1$ & $v_{11}^+ $ & $  |\zp{1}\zp{1}\rangle+\frac{a}{2}
                           (\,|\zp{3}\zn{1}\rangle+|\zn{1}\zp{3}\rangle\,)$ \\
$ 1$ & $ 1$ & $v_{12}^+ $ & $a |\zp{1}\zp{1}\rangle-
                           (\,|\zp{3}\zn{1}\rangle+|\zn{1}\zp{3}\rangle\,)$ \\
$ 1$ & $-1$ & $v_1^-    $ & $|\zp{3}\zn{1}\rangle-|\zn{1}\zp{3}\rangle$ \\
$-1$ & $ 1$ & $v_{-11}^+$ & $  |\zn{1}\zn{1}\rangle+\frac{a}{2}
                           (\,|\zn{3}\zp{1}\rangle+|\zp{1}\zn{3}\rangle\,)$ \\
$-1$ & $ 1$ & $v_{-12}^+$ & $a |\zn{1}\zn{1}\rangle-
                           (\,|\zn{3}\zp{1}\rangle+|\zp{1}\zn{3}\rangle\,)$ \\
$-1$ & $-1$ & $v_{-1}^- $ & $|\zn{3}\zp{1}\rangle-|\zp{1}\zn{3}\rangle$ \\
$ 0$ & $ 1$ & $v_{01}^+ $ & $ (\,|\zp{1}\zn{1}\rangle+|\zn{1}\zp{1}\rangle\,)+
                            b(\,|\zp{3}\zn{3}\rangle+|\zn{3}\zp{3}\rangle\,)$ \\
$ 0$ & $ 1$ & $v_{02}^+ $ & $b(\,|\zp{1}\zn{1}\rangle+|\zn{1}\zp{1}\rangle\,)-
                             (\,|\zp{3}\zn{3}\rangle+|\zn{3}\zp{3}\rangle\,)$ \\
$ 0$ & $-1$ & $v_{01}^- $ & $ (\,|\zp{1}\zn{1}\rangle-|\zn{1}\zp{1}\rangle\,)+
                            c(\,|\zp{3}\zn{3}\rangle-|\zn{3}\zp{3}\rangle\,)$ \\
$ 0$ & $-1$ & $v_{02}^- $ & $c(\,|\zp{1}\zn{1}\rangle-|\zn{1}\zp{1}\rangle\,)-
                             (\,|\zp{3}\zn{3}\rangle-|\zn{3}\zp{3}\rangle\,)$ \\
\hline
\end{tabular}
\end{center}
\end{table}
States with $p=1$ are symmetric under site transposition, the others are
antisymmetric.

For convenience, we fix the local ground state energy at zero leading to a
12-parameter model. Because this includes a trivial scale, the number of
non-trivial interaction parameters is 11.

A detailed explanation of the concept of an optimum ground state can be
found in \cite{nz}. This special type of global ground state $|\Psi_0\rangle$
satisfies
\begin{equation}
\label{frm_optcond}
h_{i,j}\, |\Psi_0\rangle \,=\, 0 \quad\mbox{for all nearest neighbours } i,j \; ,
\end{equation}
i.e. it is also a ground state of all local Hamiltonians.
\section{Vertex state model construction}
\label{sec_constr}
Trivial optimum ground states are easily obtained by using simple tensor products of
single spin states $|s_1 s_2\rangle$ for nearest neighbours. Because of
parity invariance, this also requires $|s_2 s_1\rangle$ to be a local ground state,
which leads to the ferromagnetic state of the form
\begin{equation}
|\Psi_F \rangle = |s_1 s_1 \ldots s_1 \rangle 
\end{equation}
and its antiferromagnetic counterpart
\begin{equation}
|\Psi_{AF} \rangle = \prod_{i \in A} |s_1\rangle_i \prod_{j \in B} |s_2\rangle_j \; ,
\end{equation}
if the sets $A$ and $B$ define a bipartite decomposition of the lattice and
$s_1 \neq s_2$. In the following, we focus on more complicated ground states.
Local ground states shall never be simple tensor products of the form $|s_1 s_2\rangle$,
instead they shall consist of at least two terms
\begin{equation}
|\psi_{i,j}\rangle = |s_1 s_2\rangle + c |s_1 - 1 , s_2 + 1\rangle \; .
\end{equation}
This special form ensures local $S^z$ conservation. In order to be an optimum ground
state, the global ground state $|\Psi_0\rangle$ has to be annihilated by all local
Hamiltonians $h_{i,j}$. Therefore $|\Psi_0\rangle$ contains only local ground states,
no local excited states are involved.

The global ground state $|\Psi_0\rangle$ is constructed using vertices
\begin{equation}
\vertsinglea = |s_1(\mu_1,\mu_2,\mu_3)\rangle
\end{equation}
which map the discrete bond variables $\mu_1,\mu_2,\mu_3$ to a single spin state
$|s_1(\mu_1,\mu_2,\mu_3)\rangle$.
The concatenation of two such vertices on the hexagonal lattice is defined by taking
the tensorial product of the single spin states and summing over the shared bond
\begin{equation}
\begin{array}{l}
\vertfpair \\ 
= \sum_{\mu} |s_1(\mu_1,\mu_2,\mu)\rangle \otimes
             |s_2(\mu,\mu_3,\mu_4)\rangle \\
= |s_{12} (\mu_1,\mu_2,\mu_3,\mu_4)\rangle \; .
\end{array}
\end{equation}
The result is a two-spin state which depends on the bond variables $\mu_1,\ldots,\mu_4$.
Because this concatenation can be done for an arbitrary cluster of sites
we can use it to construct a global state $|\Psi_0\rangle$ on the complete
hexagonal lattice. $|\Psi_0\rangle$ does not depend on any bond variables since
periodic boundary conditions are imposed and thus all $\mu_i$ are
summed over. This construction is much like a standard vertex model, except
that the `weights' of the vertices are single-spin states instead of numbers
and the generic product of numbers is replaced by the tensor product of spin
states. We therefore call it a {\em vertex state model}.

In order to be an optimum ground state, the constructed global state $|\Psi_0\rangle$
has to be annihilated by all local Hamiltonians $h_{i,j}$ (\ref{frm_optcond}). Since
$h_{i,j}$ only acts on the lattice sites $i$ and $j$, it is certainly sufficient
to demand
\begin{equation}
\label{frm_vertcond}
h_{i,j} \left[ \vertfpairij \right] = 0
\end{equation}
for all $\mu_1,\ldots,\mu_4$ and all neighbour sites $i,j$. In words,
the local Hamiltonian should annihilate all two-spin states, that are generated
by vertex pairs. This will ensure that only local ground states enter the global
state $|\Psi_0\rangle$.

Since we are interested in the ground state of an homogeneous global Hamiltonian, we
restrict ourselves to vertices which do not depend on the position in the lattice.
A vertex will be uniquely defined by the state of its bond variables and the sublattice
it belongs to.

A particularly interesting optimum ground state is generated by the following set of
vertices
\begin{equation}
\label{frm_vertices}
\begin{array}{rcrcrcrl}
\vertrii&:&\sigma a|\zn{3}\rangle&\hspace{1cm}&\vertiil&:&\sigma a|\zn{3}\rangle \\
\vertlii&:&        |\zn{1}\rangle&            &\vertiir&:&\sigma  |\zn{1}\rangle \\
\vertroi&:&        |\zn{1}\rangle&            &\vertiol&:&\sigma  |\zn{1}\rangle \\
\vertrio&:&        |\zn{1}\rangle&            &\vertoil&:&\sigma  |\zn{1}\rangle \\
\vertroo&:&        |\zp{1}\rangle&            &\vertool&:&        |\zp{1}\rangle \\
\vertlio&:&        |\zp{1}\rangle&            &\vertoir&:&        |\zp{1}\rangle \\
\vertloi&:&        |\zp{1}\rangle&            &\vertior&:&        |\zp{1}\rangle \\
\vertloo&:&\sigma a|\zp{3}\rangle&            &\vertoor&:&       a|\zp{3}\rangle & .
   \end{array}
\end{equation}
In this case, the bond variables $\mu_i$ can only take 2 different values, which are
denoted by arrows on the bonds. $a$ is a continuous real parameter, but $\sigma$ can
only take the values $\pm 1$. These vertices generate the following 9 two-spin
states\footnote{Common factors are divided out.}
\begin{equation}
\label{frm_localgs}
\begin{array}{rcl}
|\zp{3}\zp{1}\rangle & + & \sigma |\zp{1}\zp{3}\rangle \\
|\zn{3}\zn{1}\rangle & + & \sigma |\zn{1}\zn{3}\rangle \\
|\zp{1}\zp{1}\rangle & + & a |\zp{3}\zn{1}\rangle \\
|\zp{1}\zp{1}\rangle & + & a |\zn{1}\zp{3}\rangle \\
|\zn{1}\zn{1}\rangle & + & a |\zn{3}\zp{1}\rangle \\
|\zn{1}\zn{1}\rangle & + & a |\zp{1}\zn{3}\rangle \\
|\zp{1}\zn{1}\rangle & + & \sigma a^2 |\zp{3}\zn{3}\rangle \\
|\zn{1}\zp{1}\rangle & + & \sigma a^2 |\zn{3}\zp{3}\rangle \\
|\zp{1}\zn{1}\rangle & + & \sigma |\zn{1}\zp{1}\rangle \; ,
\end{array}
\end{equation}
which have to be annihilated by $h_{i,j}$. It turns out that this can be achieved
by putting the following constraints on the parameters defined in (\ref{frm_genham})
\begin{equation}
\begin{array}{l}
\lambda_2^\sigma    = \lambda_{11}^+      = \lambda_1^-            =
\lambda_{01}^\sigma = \lambda_{02}^\sigma = \lambda_{01}^{-\sigma} = 0 \\
b = c = \sigma a^2 \\
\lambda_3,\;\lambda_2^{-\sigma},\;\lambda_{12}^+,\;\lambda_{02}^{-\sigma} > 0 \\
a \; \mbox{arbitrary,} \; \sigma=\pm 1 \; .
\end{array}
\end{equation}
Under these conditions, the Hamiltonian still contains 5 continuous and 1 discrete
parameter ($\sigma$), but the constructed state only depends on $a$ and $\sigma$.
We like to point out that the local degeneracy, i.e. the number of local ground
states, is 9, but the global ground state is unique in the sense that it is the
only ground state of the global Hamiltonian (\ref{frm_globham}) for any finite
number of lattice sites. Appendix A contains a rigorous proof of uniqueness on
the finite hexagonal lattice. This means that condition (\ref{frm_vertcond}) is
not only sufficient to obtain an optimum ground state, but is also necessary.
\section{Properties of the state}
\label{sec_prop}
\subsection{General properties}
\label{sec_prop_gen}
By construction, the global ground state is an eigenstate of $S^z_{total}$.
In addition, it is invariant under a global spin flip $S^z \to -S^z$.
Therefore, the single spin magnetization $\langle S^z_i\rangle$ as well as the global
magnetization $\langle S^z_{total}\rangle$ vanish, which indicates an antiferromagnet.

In the special limiting case $a \to \infty $, only 4 of the 16 vertices contribute
to the global state:
\begin{equation}
\begin{array}{rcrcrcrl}
\vertrii&:&\sigma a|\zn{3}\rangle&\hspace{1cm}&\vertiil&:&\sigma a|\zn{3}\rangle \\
\vertloo&:&\sigma a|\zp{3}\rangle&            &\vertoor&:&       a|\zp{3}\rangle & .
\end{array}
\end{equation}
With these 4 vertices there are only 2 possibilities of filling the complete
lattice: A $|\zp{3}\rangle$ on the first sublattice and a $|\zn{3}\rangle$ on the
second one and vice versa. The global ground state is therefore a superposition
of both possible \neel\ states\footnote{The overall factor $a^N$ is divided out.}
\begin{equation}
\label{frm_neelcontr}
|\,\mbox{\neel}_1\,\rangle + \sigma^{N/2}\; |\,\mbox{\neel}_2\,\rangle \; .
\end{equation}
In the next subsection we present numerical investigations which indicate a
\neel\ type behaviour not only in the limit $a\to\infty$ but also for finite
values of $a$.

Another very important special case is $a=-\sqrt{3},\;\sigma=-1$. At this
{\em isotropic point} the constructed optimum ground state coincides with
the so called {\em valence bond solid} (VBS) ground state presented in
\cite{aklt,klt}. The local eigenstates are simultaneously eigenstates of
$({\bf S}_i + {\bf S}_j)^2$, so the local Hamiltonian can be chosen to have
complete $SO(3)$ symmetry
\begin{equation}
h_{i,j} =            {\bf S}_i \cdot {\bf S}_j 
  + \frac{116}{243}( {\bf S}_i \cdot {\bf S}_j )^2
  +  \frac{16}{243}( {\bf S}_i \cdot {\bf S}_j )^3 \; ,
\end{equation}
which is simply the projector onto all states with $( {\bf S}_i + {\bf S}_j )^2 = 3(3+1)$.
The authors show in \cite{aklt} that the global state has exponentially decaying
correlation functions and conjecture the existence of an energy gap. These results
are consistent with our numerical investigations.
\subsection{Transformation to a classical vertex model}
\label{sec_prop_trans}
In order to obtain correlation functions and other expectation values of the
constructed optimum ground state, we have to investigate the inner product,
i.e. the square of its norm. Sticking to the graphical language, the inner product
consists of {\em two} hexagonal lattices placed on top of each other, one for the
bra- and one for the ket-vector. As a
vertex at a specified site only generates single-spin states for this particular
lattice site, only vertex pairs at corresponding sites\footnote{
i.e. those that are directly on top of each other}
contribute to the inner product. So we just have to take the inner product at
corresponding lattice sites. The resulting model is a {\em classical} vertex
model\footnote{i.e. the vertex weights are numbers} with 6 emanating bonds as
shown in Figure \ref{fig_inner}.
\begin{figure}[p]
\caption{Two vertices at corresponding sites combine to a classical vertex
         with 6 bonds via the inner product in spin space.}
\label{fig_inner}
\end{figure}
The inner product of the global ground state is obtained by summing over all
bond configurations of this classical vertex model.

As the inner product of two different $S^z$ eigenstates is zero, only
20 of all 64 classical vertices have a non-vanishing weight. The rule that
selects these 20 vertices is the following: The number of outgoing arrows
for the bra- and the ket-vector lattice must be equal. There are two special
vertices in this classical 20-vertex model. The vertex with all 6 arrows
pointing outwards and that with all 6 arrows pointing inwards carry a weight
of $a^2$. These vertices are generated by the inner product of two
$S^z=\frac{3}{2}$ or two $S^z=-\frac{3}{2}$ vertices, respectively. All other
non-vanishing vertices carry a weight of 1. So in the special case $a=1$ the
weight of {\em all} non-vanishing vertices is 1, which corresponds to an
infinite temperature in the classical model with the usual interpretation of
a weight as `Boltzmann weight'. We shall see later that
this point lies deeply inside the disordered phase of the model.
\subsection{Numerical results}
\label{sec_prop_num}
Even for special values of the parameter $a$ no exact solution of the classical
vertex model is known. So we have to fall back on numerical methods in order
to calculate properties of the constructed global ground state. As all vertex
weights are non-negative, they can be interpreted as Boltzmann weights of a
thermodynamical system which can be handled using a Monte-Carlo-Algorithm (MCA).
According to our simulations the system exhibits a phase
transition at $a^2_c = 6.46 \pm 0.03$. For $a<a_c$ the $S^zS^z$-correlation function decays
exponentially with a correlation length of order 1. Hence the global ground state
is {\em disordered}. This phase corresponds to the Haldane phase of one-dimensional
integer spin chains. Both the isotropic point ($a^2 = 3$) and the infinite temperature
point ($a=1$) lie in this phase. But in the case of $a>a_c$ the system is dominated by
the \neel\ contributions (\ref{frm_neelcontr}) and thus the longitudinal correlation
function shows an alternating long-range behaviour. Figure \ref{fig_500800} shows
a typical correlation function for each phase. We have restricted the plot to
only one sublattice in order to suppress the effect of the alternating sign.
\begin{figure}[p]
\caption{Longitudinal correlation function in the disordered (dashed curve)
and in the \neel\ phase (full curve).}
\label{fig_500800}
\end{figure}

To determine the type of phase transition, we have investigated the {\em probability
distribution of Boltzmann weights} directly at $a_c$. In our case, there is only
{\em one} peak in this distribution, which clearly indicates a second order phase
transition. For a first order transition, the phase coexistence would yield at
least two peaks in the probability distribution \cite{clb}.

Consistently, the longitudinal correlation function decays algebraically at the
critical point. Fig. \ref{fig_644} shows the measured correlation function for a
\begin{figure}[p]
\caption{Longitudinal correlation function at the critical point (dots) and
the fitted algebraic function (\ref{frm_fit}).}
\label{fig_644}
\end{figure}
system with $2\times 60\times 60$ sites. Again, we have restricted ourselves to only
one sublattice to avoid the alternation effect. The figure also shows the function
\begin{equation}
\label{frm_fit}
f_l(r)=c_l \left(\frac{1}{\sqrt{r}} + \frac{1}{\sqrt{61-r}} \right) \; ,
\end{equation}
which connects the measured points nearly perfectly if $c_l$ is fitted to
$1.62 \pm 0.01$. The second term in (\ref{frm_fit}) symmetrizes the function,
which compensates the effect of periodic boundary conditions in first order.
\section{Asymptotic equivalence to a free-fermion model}
\label{sec_free}
\subsection{Diagonal vertices and transformation}
\label{sec_free_trans}
Instead of investigating further critical properties using numerical data, we will
now show that the model is asymptotically equivalent to an exactly solvable model for
most values of the parameter $a$.

The classical vertex model introduced in subsection \ref{sec_prop_trans} consists of two
different types of vertices: If all 3 arrows of a vertex in the bra-vector lattice are
equal to those in the ket-vector lattice, we call this vertex {\em diagonal}.
Conversely, {\em off-diagonal} vertices contain at least one unequal arrow pair\footnote{
Actually, these vertices contain exactly two unequal arrow pairs, otherwise the corresponding
vertex weight is zero.}. On each sublattice, there are 8 diagonal and 12 off-diagonal
vertices with non-vanishing weights. The crucial result of the Monte-Carlo simulations
is that the probability of an unequal arrow pair on a bond decays exponentially with
increasing values of $a^2$. As shown in Fig. \ref{fig_unequal}, $p_{\downarrow\uparrow}$
is already smaller than $10^{-4}$ for $a^2=5$.
\begin{figure}[p]
\caption{Probability of unequal arrow pairs as a function of $a^2$ for a system of
$24\times 24$ sites (full curve) and $60\times 60$ sites (dashed curve).}
\label{fig_unequal}
\end{figure}
Since $a_c^2 > 6$, off-diagonal vertices are completely negligible
in the regime of the phase transition. Properties exactly at the critical point
should be independent of the model details anyway. This leads us to the diagonal
vertex model, which is of course a model with only {\em one} arrow on each bond.
This model on the hexagonal lattice consists of the vertices
\begin{equation}
\label{frm_diagvert}
\begin{array}{rclcrcl}
\vertxrii&:& a^2 &\hspace{1cm}&\vertxloo&:& a^2 \\
\vertxlii&:& 1   &            &\vertxroo&:& 1   \\
\vertxroi&:& 1   &            &\vertxlio&:& 1   \\
\vertxrio&:& 1   &            &\vertxloi&:& 1
   \end{array}
\end{equation}
and identical vertices on the other sublattice. We shall now show that this model is
a free-fermion model and thus exactly solvable.

The first step is to reinterpret the hexagonal lattice as a square lattice with an
elementary cell that consists of two neighbour vertices. Such a pair of vertices can be
combined to a single one by simply summing out the connecting bond. The result is a
16-vertex model on the square lattice with the following properties:
\begin{enumerate}
\item All 16 vertices have a non-vanishing weight.
\item It is a zero-field vertex model, i.e. each vertex is invariant under
      a flip of all arrows.
\end{enumerate}
The second property enables us to transform the model to an 8-vertex model using the
following orthogonal transformation.

Let us denote the vertex weights of the 16-vertex model by $w(i_1,i_2,i_3,i_4)$
in which the $i_{\nu}$ can take the value 1 for a left/down arrow or 2 for a right/up
arrow. We now define a new vertex model on the square lattice
\begin{equation}
\begin{array}{l}
w'(j_1,j_2,j_3,j_4)= \\
\sum_{i_1,i_2,i_3,i_4} w(i_1,i_2,i_3,i_4) \;
u_{i_1 j_1} u_{i_2 j_2} u_{i_3 j_3} u_{i_4 j_4}
\end{array}
\end{equation}
using the orthogonal $2\times 2$-matrix
\begin{equation}
u=\frac{1}{\sqrt{2}}\left(
\begin{array}{rr} 1 & 1 \\ -1 & 1 \end{array}\right) \; .
\end{equation}
If we concatenate two $w'$-vertices, both $u$-matrices on the connecting bond
cancel out, because $u$ is orthogonal. The $w'$-vertex model is therefore completely
equivalent to the original $w$-vertex model. Due to the zero-field symmetry of the
$w$-vertices, only those $w'$-vertices with an even number of outgoing arrows have
a non-vanishing weight, i.e. we are left with an 8-vertex model. If the usual convention
for vertex numbering in the 8-vertex model is used (see e.g. \cite{baxter} or \cite{fw}),
the weights of the transformed model can be summarized as
\begin{equation}
\label{frm_8vmweights}
\begin{array}{l}
\omega_1=\frac{1}{2} (3+a^2)^2 \\
\omega_2=\omega_3=\omega_4=\omega_5=\omega_6=\frac{1}{2}(-1+a^2)^2 \\
\omega_7=\omega_8=\frac{1}{2}(1+a^2)^2 - 2 \; .
\end{array}
\end{equation}
Fortunately, these weights fulfil the {\em free-fermion condition}
\begin{equation}
\omega_1 \omega_2 + \omega_3 \omega_4 =
\omega_5 \omega_6 + \omega_7 \omega_8
\end{equation}
for all values of $a$, which means the model is exactly solvable.
\subsection{Solution of the free-fermion model}
\label{sec_free_solution}
A detailed investigation and solution of free-fermion models has been done by {\sc Fan}
and {\sc Wu} in \cite{fw} using the method of dimers. Applying their general solution
to our specific model, the partition function is given by
\begin{equation}
\ln Z=\frac{1}{4\pi} \int_0^{2\pi} \ln \left(\,A(\phi) + \sqrt{Q(\phi)}\,\right) \;d\phi \; ,
\end{equation}
in which
\begin{equation}
\label{frm_apqp}
\begin{array}{l}
A(\phi) = \\
\frac{1}{2}(a^8 + 18 a^4 + 24 a^2 + 21) + 2(a^4 - 1)(a^2 - 1) \cos \phi \\
Q(\phi) = \\
\left( 2(a^4 - 1)(a^2 - 1) \cos \phi - \frac{1}{2}((a^2 + 1)^2 + 4)^2 \right)^2 - \\
2(a^4 + 3)(a^2 + 1)((a^2 + 1)^2 - 4)^2 \; .
\end{array}
\end{equation}

We wish to consider two special points. The first one is of course $a^2=1$. As stated
earlier, this corresponds to infinite temperature, because all weights of the
original $w$-model are equal to 1, i.e. disorder is maximized. In this context, we
have to clearify the role of the orthogonal transformation $u$ as a kind of {\em duality
transformation}. At $a^2=1$ only $\omega_1$ out of the vertex weights (\ref{frm_8vmweights})
of the $w'$-model is non-vanishing, so in the language of the $w'$-model, temperature is
exactly zero. Conversely, the limit $a^2 \rightarrow \infty$ corresponds to $T=0$ in the
original $w$-model and to $T \rightarrow \infty$ in the $w'$-model. So the $u$
transformation exchanges high- and low-temperature limits. Of course, the structure of the
physical quantum spin ground state is only reflected by the original model, since
the single spin state at a given site is determined by the number of outgoing arrows of
the corresponding $w$-vertex.

The second important point is defined by $Q(\phi)=0$,
where the system exhibits a second order phase transition as explained in \cite{fw}.
Solving this condition for $a^2$ yields
\begin{equation}
a_c^2 = 3+\sqrt{12} = 6.464 \ldots
\end{equation}
which is in perfect agreement with our numerical result.

A note added in proof is contained in \cite{fw}, which states that this free-fermion
model is actually equivalent to the Ising model on an anisotropic triangular lattice.
This equivalence is capable of explaining the exponent $\eta=\frac{1}{2}$ in the decay
of the critical correlation function (\ref{frm_fit}). Generally, the correspondence
between arrow variables $a_i$ of the vertex model and Ising spin variables
$\sigma_{i'}$ is
\begin{equation}
a_i = \sigma_{i'} \, \sigma_{i'+1} \; ,
\end{equation}
in which $i'$ and $i'+1$ are the sites on the Ising lattice that `enclose' the vertex
model bond $i$. We can now express correlation functions of the vertex model in terms
of Ising model correlations
\begin{equation}
\langle a_0\,a_r\rangle = \langle \sigma_0\,\sigma_1\,\sigma_{r'}\,\sigma_{r'+1}
\rangle \; .
\end{equation}
The leading contribution to this 4-spin correlation function should be
\begin{equation}
\langle \sigma_0\,\sigma_1\,\sigma_r\,\sigma_{r+1} \rangle
\simeq \langle \sigma_0\,\sigma_r \rangle^2 \propto r^{-\frac{1}{2}}
\; ,
\end{equation}
since the critical 2-spin correlations of the Ising model decay as $r^{-\frac{1}{4}}$.
This result is completely consistent with equation (\ref{frm_fit}), which is a
surprisingly good fit to the simulated critical correlation function.
\section{Conclusion}
\label{sec_concl}
We have constructed a one-parametric set of optimum ground states for a five-dimensional
manifold of spin-$\frac{3}{2}$ Hamiltonians with nearest-neighbour interaction on the
hexagonal lattice.
These ground states are antiferromagnetic in the sense that the global magnetization
vanishes for all values of the continuous parameter $a$. Uniqueness of the ground states
on finite lattices with periodic boundary conditions can be proven rigorously using
an induction technique. For special values of the parameters, this ground state coincides
with the VBS ground state introduced in \cite{aklt,klt}.

In order to investigate the properties of the constructed ground state we have implemented
a Monte-Carlo algorithm. The result of the simulations is that there is a
parameter-induced phase transition from a disordered phase with exponentially
decaying correlation functions to a \neel\ ordered phase with long-range correlations.
This phase transition at $a^2_c\approx 6.46$ is of second order. The longitudinal
two-spin correlation function decays algebraically at this point with an exponent of
$\eta=\frac{1}{2}$.

The norm of the constructed global ground state can be regarded as the partition function
of a classical
vertex model on the hexagonal lattice with 4 bond states. One result of our numerical
simulations is that this 4 bond state model can be reduced to a free-fermion
model with only 2 bond states in good approximation for $a^2>2$. The exact solution
of the free-fermion model reproduces our numerical results with high accuracy.
\appendix
\renewcommand{\thesection}{Appendix \Alph{section}:}
\renewcommand{\theequation}{\Alph{section}.\arabic{equation}}
\setcounter{equation}{0}
\section{Proof of uniqueness}
\label{sec_proof}
To prove the uniqueness of the optimum ground state that we constructed in section
\ref{sec_constr} on the finite lattice, we start with an `empty' hexagonal lattice
with no spins at the
sites. Now we gradually fill this lattice by adding one vertex after another (as
defined in (\ref{frm_vertices})) at arbitrary empty sites. Whenever two neighbouring
sites are occupied (periodic boundary conditions are imposed), we sum over the arrow
configurations of the bond between them. After having added several vertices and the
lattice is partially filled, we have two types of bonds: {\em Internal bonds} between
two occupied sites, where we have summed over, and {\em external bonds} between an
occupied and an unoccupied site with a variable arrow on it. During the process
of filling, the lattice might exhibit arbitrarily shaped and possibly unconnected
clusters. Finally, after occupying all sites, we have summed over all bonds
and end up with exactly the desired global optimum ground state.

This step by step construction enables us to prove the uniqueness of the global
ground state by induction according to system size.
After each step, i.e. every time we have added one vertex,
we prove the following induction hypothesis:
\begin{quote}
Running through all arrow configurations on the external bonds
yields a set of global ground states, which generates {\em all}
global ground states of the current spin distribution. \\
Linear dependencies are only introduced by topologically
equivalent external bonds.\footnote{We call external bonds
topologically equivalent if they are emanated by the same
lattice site. A pair of unequal arrows on 2 topologically
equivalent external bonds can be swapped without changing
the global state.}
\end{quote}
Since there are no external bonds on the completely filled
lattice, we shall have proven uniqueness of the constructed optimum
ground state.

For the very first step it is immediately clear from the list
of vertices (\ref{frm_vertices}) that the induction hypothesis
holds, because we can realize all single-spin states on both
sublattices.

Now we assume that the hypothesis holds for systems of size
$N$ (and all smaller sizes) and add a single vertex. We have
to consider 4 different cases:
\begin{enumerate}
\item The new vertex is isolated, i.e. it has no occupied
      neighbours.

In this case the hypothesis is trivial. There is no local
Hamiltonian that connects the new spin with any other spin.
So the ground states of the $N+1$ spin system are simply all
tensorial products of $N$ spin ground states and all single-spin
states. Because we can obviously generate all single-spin states
with an isolated vertex, we are immediately done by using the
assumption for $N$ spins. Degeneracy is also trivial.
\item The new vertex has exactly one occupied neighbour.

Let $i_0$ denote the site where the new vertex has been added
and $i_1$ the site of its neighbour. Now a new local Hamiltonian
$h_{i_0,i_1}$ comes into play, which has to annihilate the
global ground state. In order to prove the induction hypothesis
for the $N+1$ spin system, we split it into 3 parts:
The vertex $|\phi\rangle_{i_0}^{\alpha}$         at site $i_0$,
the vertex $|\phi\rangle_{i_1}^{\beta_1,\gamma}$ at site $i_1$,
and the remaining vertex state model for $N-1$ spins
$|\Phi\rangle^{\beta_2,\gamma}$.
In this notation, the ground state condition is
\begin{equation}
\label{frm_proof_ansatz}
h_{i_0,i_1} \; \sum_{\alpha,\beta_1,\beta_2,\gamma}
A_{\alpha}^{\beta_1,\beta_2}
|\phi\rangle_{i_0}^{\alpha}
|\phi\rangle_{i_1}^{\beta_1,\gamma}
|\Phi\rangle^{\beta_2,\gamma} = 0
\end{equation}
The meaning of the indices is the following. $\alpha$ represents
all bonds of the new vertex at $i_0$. $\beta_1$ are the external
bonds of the vertex at $i_1$ {\em before the new vertex has been
added}. $\beta_2$ are the external bonds of the remaining vertex
state model. $\gamma$ represents the internal bonds between $i_1$
and the $N-1$ spin system. $A_{\alpha}^{\beta_1,\beta_2}$ is a
tensor of numbers which has to be determined.

Because $h_{i_0,i_1}$ only acts on $i_0$ and $i_1$, the idea is
to eliminate $|\Phi\rangle^{\beta_2,\gamma}$ from
(\ref{frm_proof_ansatz}). Normally, we can only do that if the
states $|\Phi\rangle^{\beta_2,\gamma}$ constitute a basis of $N-1$
spin space, in particular, they have to be linearly independent.
In our case, we can still drop $|\Phi\rangle^{\beta_2,\gamma}$,
because of the following reason:

The linear dependencies can be eliminated just by restricting
the values of the $\beta_2$ index, but leaving the $\gamma$
bonds untouched.
Figure \ref{fig_neighbours} shows all neighbours of a
new added vertex. These neighbours are all linked to 6
{\em different} sites. So each vertex never emanates more
than one $\gamma$-bond.\footnote{This is always true on a
bipartite lattice.} Since linear dependencies can only enter
through topologically equivalent external bonds
(induction hypothesis), only the
$\beta_2$ bonds have to be restricted. 
We can therefore exclude all over-counted states by
restricting the $A_{\alpha}^{\beta_1,\beta_2}$-tensor,
which means no loss in generality, because we are only
interested in states, not representations.
\begin{figure}[p]
\caption{A newly added vertex (unfilled circle) and its neighbours.
           Note that all neighbours are linked to different sites of
           the remaining lattice. So the bonds between the neighbours
           and the remaining lattice (thick lines) can never cause
           linear dependencies.}
\label{fig_neighbours}
\end{figure}

After dropping the $|\Phi\rangle^{\beta_2,\gamma}$, we are left with
the following condition
\begin{equation}
\label{frm_proof_simple}
h_{i_0,i_1} \; \sum_{\alpha,\beta_1}
A_{\alpha}^{\beta_1,\beta_2}
|\phi\rangle_{i_0}^{\alpha}
|\phi\rangle_{i_1}^{\beta_1,\gamma} = 0
\mbox{, for all $\beta_2,\gamma$ ,}
\end{equation}
which is nothing but the ground state condition for the two-spin
system. Of course, this condition yields exactly the two-spin states
(\ref{frm_localgs}) as its solutions, because the local Hamiltonian $h_{i_0,i_1}$
is constructed to annihilate exactly these 9 local states.
As our vertex state model generates these 9 two-spin states,
the first part of the induction hypothesis is proven.

To check the statement about linear dependency we decompose the
system into two parts: The two-spin system $i_0,i_1$ and the remaining
$N-1$ spin system. We have already proven that the interface between
these two parts -- the $\gamma$ bonds -- cannot introduce linear
dependencies. So the only remaining sources are the external bonds of
the whole $N+1$ spin system, which immediately yields the induction
hypothesis.

The above considerations are only valid for $a\neq0$ and $a\neq\infty$,
otherwise equation (\ref{frm_proof_simple}) has more solutions than
two-spin states are generated by vertex pairs. Consequently the global
ground state is highly degenerate if $a=0$ or $a\to\infty$.
\item The new vertex has two occupied neighbours.

In this case, the situation is only slightly more complicated than in the
previous case. Instead of only one neighbour $i_1$ we now have $i_1$ and
$i_2$. Consequently, $h_{i_0,i_1}$ and $h_{i_0,i_2}$ are the new local
Hamiltonians that have to annihilate the global state\footnote{
$i_1$ and $i_2$ can never be crosslinked, because they belong to the same
sublattice.}.
The ground state condition is
\begin{equation}
\begin{array}{l}
\left( h_{i_0,i_1} + h_{i_0,i_2} \right) \;
\sum_{\alpha,\beta_1,\beta_2,\beta_3,\gamma_1,\gamma_2}
A_{\alpha}^{\beta_1,\beta_2,\beta_3} \;\cdot \\
\quad
|\phi\rangle_{i_0}^{\alpha}
|\phi\rangle_{i_1}^{\beta_1,\gamma_1}
|\phi\rangle_{i_2}^{\beta_2,\gamma_2}
|\Phi\rangle^{\beta_3,\gamma_1,\gamma_2} = 0 \; .
\end{array}
\end{equation}
Since the argument of the previous case holds for any number of
neighbours, we can eliminate
$|\Phi\rangle^{\beta_3,\gamma_1,\gamma_2}$ and are left with
\begin{equation}
\begin{array}{l}
\left( h_{i_0,i_1} + h_{i_0,i_2} \right) \;
\sum_{\alpha,\beta_1,\beta_2}
A_{\alpha}^{\beta_1,\beta_2,\beta_3} \;\cdot \\
\quad
|\phi\rangle_{i_0}^{\alpha}
|\phi\rangle_{i_1}^{\beta_1,\gamma_1}
|\phi\rangle_{i_2}^{\beta_2,\gamma_2} = 0
\mbox{, for all $\beta_3,\gamma_1,\gamma_2$ ,}
\end{array}
\end{equation}
which is the ground state condition for a 3-spin system. But this system
can be constructed using only case 1, so we have already proven that we
can generate all ground states of this 3-spin system using our vertex state
model.

Balance of linear dependencies is checked exactly the same way as in case 1.
Our induction hypothesis is therefore also valid in the case of two
occupied neighbours.
\item The new vertex has only occupied neighbours.

Because the argument is completely analogous to case 2, we omit the
corresponding ground state conditions here. We end up with a 4-spin system,
which can be handled completely by using case 1.
\end{enumerate}
We are now able to generate all ground states for every spin distribution on
the finite hexagonal lattice with periodic boundary conditions by running
through all arrow configurations on the external bonds. Because the completely
occupied lattice does not contain any external bonds, we have proven
uniqueness of the global ground state that we constructed in section \ref{sec_constr}
on the finite hexagonal lattice.


\begin{thebibliography}{9}
\bibitem{ksz1}   Kl\"{u}mper, A., Schadschneider, A., Zittartz, J.:
                 Z. Phys. {\bf B87}, 281 (1992)
\bibitem{ksz2}   Kl\"{u}mper, A., Schadschneider, A., Zittartz, J.:
                 Europhys. Lett. {\bf 24}, 293 (1993)
\bibitem{ksz3}   Kl\"{u}mper, A., Schadschneider, A., Zittartz, J.:
                 J. Phys. {\bf A24}, L955 (1991)
\bibitem{nz}     Niggemann, H., Zittartz, J.:
                 Z. Phys. {\bf B101}, 289 (1996)
\bibitem{aklt}   Affleck, I., Kennedy, T., Lieb, E.H., Tasaki, H.:
                 Commun. Math. Phys. {\bf 115}, 477 (1988)
\bibitem{klt}    Kennedy, T., Lieb, E.H., Tasaki, H.:
                 J. Stat. Phys. {\bf 53}, 383 (1988)
\bibitem{clb}    Challa, M.S.S., Landau, D.P., Binder, K.:
                 Phys. Rev. {\bf B34}, 1841 (1986)
\bibitem{baxter} Baxter, R.J.: {\em Exactly Solved Models in Statistical Mechanics},
                 Academic Press (1982)
\bibitem{fw}     Fan, C., Wu, F.Y.:
                 Phys. Rev. {\bf B2}, 723 (1970)
\end{thebibliography}
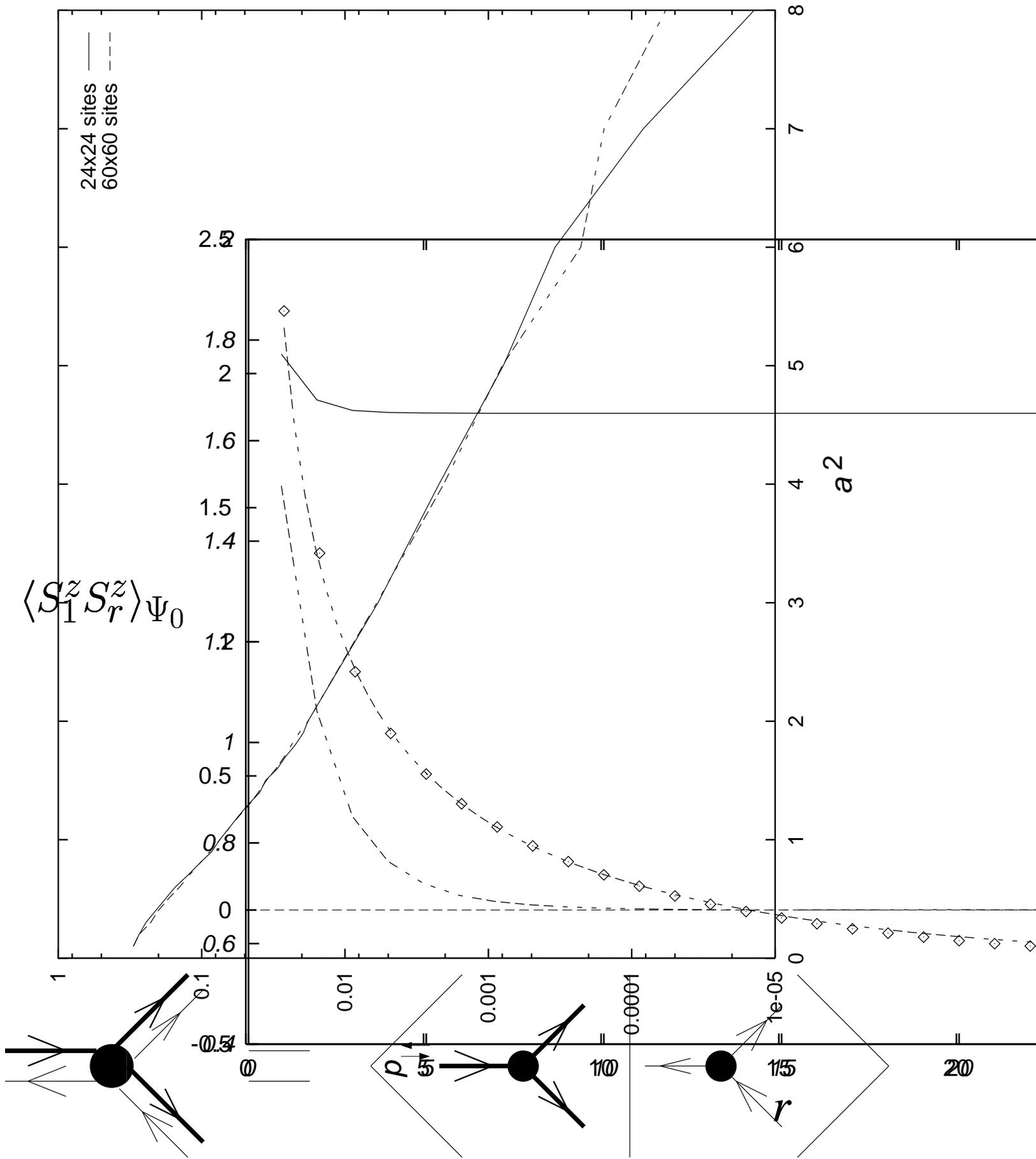
\end{document}